\documentclass[11pt]{article}\usepackage[dvips]{graphicx}
\usepackage{latexsym}
\usepackage{amssymb}
\newcommand{\R}{\mathbb{R}}
\usepackage{amsmath, amsthm, amssymb}

\def\bea{\begin{eqnarray}}
\def\eea{\end{eqnarray}}

\def\be{\begin{equation}}
\def\ee{\end{equation}}
\def\c{\check}

\begin{document}
\title{\vskip -70pt
%\begin{flushright}
%{\normalsize DAMTP-2008-58} \\
%\end{flushright}
\vskip 80pt
{\bf Cosmic Jerk and Snap in Penrose's CCC model}
\vskip 20pt}
\author{Maciej Dunajski\thanks{\tt m.dunajski@damtp.cam.ac.uk}
\\[8pt]
{\sl Department of Applied Mathematics and Theoretical Physics} \\
{\sl University of Cambridge} \\
{\sl Wilberforce Road, Cambridge CB3 0WA, UK} \\[8pt]}
\date{} 
\maketitle
\begin{abstract} 
We obtain a constraint on cosmological scalars for the FRW
metric with a pure radiation fluid source and positive cosmological constant.
We demonstrate that this constraint is conformally invariant in the context of Penrose's Conformal Cyclic Cosmology proposal, where the metrics of the late stages of the previous aeon and the early stages of the present aeon
are described by FRW cosmologies.
\end{abstract}
%\section{Introduction}
%\setcounter{equation}{0}
Conformal Cyclic Cosmology (CCC) of Penrose is a cosmological model which postulates the existence of an infinite sequence of aeons, and provides
a radical alternative to Inflationary Cosmology \cite{penrose}. The aeon
$(\c{M}, \c{g})$ (which we shall call the present aeon) starts on an initial Big-Bang singularity $\Sigma$, which 
is identified with a space-like future null infinity of the 
exponentially expanding
aeon
$(\hat{M}, \hat{g})$ (from now on called the previous aeon). 
The cosmological constant of all aeons is assumed to be positive.
The bridging space-time $M=\hat{M}\cup\c{M}\cup\Sigma$
is equipped with a regular Lorentzian metric $g$  such that
\[
\hat{g}=\hat{\Omega}^2 g, \quad \c{g}=\c{\Omega}^2 g, \quad
\mbox{and} \quad \Sigma=\{\hat{\Omega}^{-1}=0\}=\{\c{\Omega}=0\}.
\]
The conformal factors satisfy the reciprocal hypothesis $\hat{\Omega}=
-\c{\Omega}^{-1}$, and are determined by a Yamabe-type equation on the
$(M, g)$ background.
The Big-Bang three-surface $\Sigma$ is singular in the present aeon, but
this singularity is only manifest in the conformal factor $\c{\Omega}$.
The Weyl curvatures of $g, \hat{g}, \c{g}$ are all equal and are assumed to be finite at  $\Sigma$. Penrose argues that this is in agreement with the second
law of thermodynamics which requires the initial gravitational entropy 
to be low. This is to be contrasted with a final black hole singularity 
(of say the Schwarzchild solution) where the Weyl tensor blows up.

We shall assume that the past aeon was described by an 
FRW metric 
\be
\label{FLRW}
\hat{g}=-d\hat{t}^2+\hat{a}^2h
\ee
with a pure radiation fluid source and positive cosmological constant. Here  $\hat{a}=\hat{a}(\hat{t})$ 
is the scale factor, and
$h$ is a metric on $H^3$, $\R^3$ or $S^3$ with constant 
curvature $\hat{k}=-1, 0$ or $1$. To the first order in time, the 
scale factor of this aeon is measured by
the Hubble scalar with a dimension of inverse of time. 
The next three
terms in the Taylor expansion
of the red--shift are given by dimensionless scalars
called 
deceleration, jerk and snap respectively
\be
\label{scalars}
\hat{q}=-\hat{a}
\Big(\frac{d \hat{a}}{d \hat{t}}\Big)^{-2}\frac{d^2 \hat{a}}{d \hat{t}^2}, \quad
\hat{Q}=\hat{a}^2
\Big(\frac{d \hat{a}}{d \hat{t}}\Big)^{-3}\frac{d^3 \hat{a}}{d \hat{t}^3}, 
\quad 
\hat{X}=\hat{a}^3
\Big(\frac{d \hat{a}}{d \hat{t}}\Big)^{-4}\frac{d^4 \hat{a}}{d \hat{t}^4}.
%\quad 
%Y=a^4
%\Big(\frac{d a}{d t}\Big)^{-5}\frac{d^5 a}{d t^5}.
\ee
%We shall assume that the past aeon was described by an 
%FLRW metric 
%\[
%\hat{g}=-d\hat{t}^2+\hat{a}^2h
%\]
The Einstein equations with cosmological constant $\hat{\Lambda}$
reduce to the Friedmann equation
\be
\label{Friedmann}
\Big(\frac{d \hat{a}}{d \hat{t}}\Big)^2+\hat{k}=\frac{8\pi G}{3}\hat{\rho} 
\hat{a}^2+\frac{\hat{\Lambda}}{3}\hat{a}^2,
\ee
with the
energy-momentum 
conservation equations giving rise  to the density $\hat{\rho}=m\hat{a}^{-4}$.

The Friedmann equation can be reinterpreted as 
an algebraic constraint between the scalars (\ref{scalars}). 
The general procedure
for obtaining such constraints for any matter model has been described in
\cite{DG08}, and some special cases were discussed in  \cite{Harrison, Visser}.
This links the measurement
of the scalars (\ref{scalars}) to a test of General Relativity,
or any of its modifications
in the spirit of \cite{statefinder_cite}.
If one {assumes} that Einstein equations hold, then measuring
the cosmological scalars could determine the equation of state relating
the energy density and the momentum in the perfect fluid energy momentum 
tensor.

To derive the constraint consider a system of three equations
consisting of (\ref{Friedmann}) and its first two time derivatives.
We regard this as a system of algebraic equations for
the constants $(\hat{k}, \hat{\Lambda}, m)$ which can therefore be expressed
as functions of $(\hat{a}, \dot{\hat{a}}, \ddot{\hat{a}}, \dddot{\hat{a}})$. 
Take
the third derivative of (\ref{Friedmann}) and substitute the
expressions for $(\hat{k}, \hat{\Lambda}, m)$. This yields
\be
\label{our_constraints}
\hat{X}+3(\hat{q}+\hat{Q})+\hat{q}\hat{Q}=0.
\ee
This fourth order ODE is equivalent to the Friedmann equation
and has an advantage that it appears as a constraint on directly
measurable quantities (\ref{scalars}). 

%If only two constants
%$(\hat{\Lambda}, GM)$ are eliminated between (\ref{Friedmann}) and its first
%derivative then the second derivative of  (\ref{Friedmann}) yields
%\[
%2\hat{k}=\hat{a}^2\hat{H}^2(\hat{Q}-\hat{q}-2)
%\]
%obtained in \cite{Harrison}. In particular if $\hat{k}=0$ this relation reduces
%to a  third order ODE relating the jerk and the deceleration
%\[
%\hat{Q}-\hat{q}=2.
%\]
%\subsection{Conformal transformations}
%The general form of an $n$th cosmic scalar is
%\[
%Q^{(n)}=a^{n-1}\Big(\frac{da}{dt}\Big)^{-n}\frac{d^n a}{dt^n}.
%\]
%Consider a time-dependent conformal rescalling of the metric (\ref{FLRW})
%\[
%\hat{g}=\Omega^{2}(t)g.
%\]
%This metric is also of the FLRW form, and its cosmic scalars are given by
%\[
%\hat{Q}^{n}=\frac{\Omega^{n-1}a^{n-1}}{(a\dot{\Omega}/\Omega+\dot{a})^n}
%\Big(\frac{1}{\Omega}\frac{d}{dt}\Big)^n(\Omega a).
%\]
According to Tod \cite{tod2}, the metric $\c{g}$ in the present aeon 
is conformally
related to the FRW metric $\hat{g}$ in the past aeon
by
\[
\c{g}=\alpha\;\hat{a}^{-4}\hat{g} =-d\c{t}^2+\c{a}^2h,
\]
where $\alpha>0$ is a constant. As both metrics are of the FRW form, we can compute the conformal transformation properties of the cosmic 
deceleration, jerk and snap respectively:
\[
\c{q}=-\hat{q},\quad \c{Q}=\hat{Q}-2\hat{q}, \quad \c{X}=-\hat{X}-6\hat{Q}+6\hat{q}-2\hat{q}^2.
\]
Therefore the cosmological scalars are not conformally invariant.
We nevertheless explicitly verify that the constraint (\ref{our_constraints})
is conformally invariant up to an overall sign: it holds both in the past and the current aeons.

Our findings agree with the calculation of Tod \cite{tod2} and Newman \cite{newman},  who showed that both aeons are diffeomorphic if either of them is described by pure radiation FRW cosmology. In \cite{tod2} Tod uses the FRW example to motivate a general prescription
of finding a conformal factor relating two non-conformally flat aeons,
where the past aeon described by a Bianchi type cosmological model. It
would be  interesting to extend our procedure to this case, where the
cosmological scalars (\ref{scalars}) presumably need to be constructed out
of the total volume element  of the spatial slice.

\end{document}